\documentclass[twocolumn,aps,prd,floats,floatfix,showpacs,superscriptaddress,nofootinbib]{revtex4}
\usepackage{amsmath,amsfonts}
\usepackage{graphicx}

\begin{document}

\title{ Bondi accretion onto cosmological  black holes  }
\author{Janusz Karkowski}
\author{Edward Malec}
\affiliation{Instytut Fizyki Mariana  Smoluchowskiego,  Instytut Fizyki, Uniwersytet Jagiello\'nski,Reymonta 4, 30-059 Krak\'{o}w, Poland }

\begin{abstract} 
In this paper we investigate  a steady accretion within the Einstein-Straus vacuole, in the presence of the cosmological constant. The dark energy damps the mass accretion rate and --- above certain limit --- completely stops the steady accretion onto  black holes, which in particular  is prohibited   in the inflation era and  after (roughly) $10^{12}$ years from Big Bang (assuming the presently known value of the cosmological constant).  Steady accretion would not exist in the late phases of the  Penrose's scenario -  known as the   Weyl curvature hypothesis  -      of the evolution of the Universe.
\end{abstract}

\maketitle 

\section{Introduction}

The classical Bondi accretion model \cite{Bondi}  describes, in the test gas and stationarity approximations, the spherical accretion/wind of a barotropic fluid onto a newtonian gravity center. That description has been later extended to general-relativistic spacetimes.   The steady accretion in the   Schwarzschildean geometry has been investigated by Michel \cite{Michel} and Shapiro and Teukolsky \cite{shapiro_teukolsky}, and  in a spherical symmetric spacetime with backreaction  by Malec \cite{Malec99}.  A remarkable universal behaviour has been discovered  for selfgravitating transonic flows, under suitable boundary conditions: the ratio of respective mass accretion rates $\dot M_{GR} /\dot M_N$ appears  independent of the fractional mass of the gas and depends only on the asymptotic temperature. It is close to 1 in the regime of low asymptotic temperatures and can grow by one order of magnitude at high temperatures \cite{Malec2006}-\cite{Rembiasz2010}.  Here $\dot M_{GR}$  and  $\dot M_N$  are  the general relativistic and the Newtonian mass accretion rates, respectively.
The stability of steady accretion has been established  by Mach and his coworkers \cite{Mach_stability1} --- \cite{Mach_stability2}.

In this paper we consider the accretion onto a black hole that is immersed in a cosmological universe.
We adopt the Einstein-Straus "swiss cheese" model \cite{Einstein1945}. A spherically symmetric black hole surrounded with accreting gas sits inside a vacuole that is matched --- through  the  Darmois-Israel   junction conditions \cite{Darmois}, \cite{Israel} ---    to  an external FRLW  spacetime. We calculate how the presence of   a cosmological constant $\Lambda $ would affect  the most important system characteristic --- the mass accretion rate $\dot M$ of a fluid onto the center. We found that the mass accretion rate  is damped by the presence of the cosmological constant. This effect is stronger for negative values  than for   positive values of the  cosmological constant and it depends on the ratio of the dark energy density to the fluid density.  

In the next two sections we remind the Kottler solutions and draw the general picture of the steady accretion, together with  suitable equations. Sections 4 and 5 present three analytic sets of results. These show that the matter distribution can be affected by the cosmological system. More importantly, we prove that if the  cosmological constant exceeds certain limit, then the steady accretion ceases to exist.
Section 6 presents results of numerical integrations. It appears that the presence of dark energy, either negative (attractive) or positive (repulsive), damps the mass accretion rate.  It appears - in accordance with analytic predictions -  that the steady accretion does not exist above certain values of the cosmological constant. 
The last section summarizes main results and   applies them to two cosmological epochs when the dark energy is dominant. It is pointed out that our results do not agree with the Penrose's " Weyl curvature hypothesis".

\section{Accretion in an Einstein-Straus vacuole }

 We assume relativistic units with $G=c=1$. Let a FLRW spacetime be filled with dust and dark energy. Its line element reads 
\begin{equation}
ds^2 = -  d\tau^2   +a^2(\tau ) \left( {dr^2\over   1-  kr ^2} +r^2 d\Omega^2\right) , 
\label{FLRW}
\end{equation}
where the variables have obvious meaning \cite{Weinberg08}  and $k=0, ^+_-1$.  We do not need to analyze the FLRW equations, for a reason explained below.
 
We cut off, following Einstein and Straus \cite{Einstein1945},   a ball    from this   FLRW spacetime and insert instead a spherical symmetric inhomogeneity  --- a black hole surrounded by a spherical cloud of gas ---   satisfying Einstein’s equation. It is known that the boundary of the ball must be comoving if the FLRW geometry contains only dust and the dark energy represented by $\Lambda $, assuming the absence of a boundary layer  and continuity of pressure \cite{Lake2011}, \cite{Maartens2012}, \cite{Giulini2010}.  Crenon and Lake show that the no-boundary layer assumption is necessary for having a comoving boundary  \cite{Lake2011}.

Inside the excision  ball itself we can distinguish an inner accretion region such that the areal radius $R$ does not exceed a limiting radius $R_\infty $ (that in turn  should be  smaller, or perhaps much smaller, than the areal radius of the boundary of the ball),   and the  asymptotic vacuum region    outside  $R_\infty $.  In the asymptotic region the metric is given by the Kottler   ( Schwarzschild-de Sitter (SdS,  $\Lambda >0$)  and Schwarzschild-anti de Sitter (SAdS, $ \Lambda <0$) ) spacetime line element \cite{Perlick2004}
\begin{equation}
ds^2 = -\left( 1-{2m\over R}-{\Lambda \over 3} R^2\right) dt^2   +{dR^2\over   1-{2m\over R}-{\Lambda \over 3} R^2} +R^2 d\Omega^2.
\label{sds}
\end{equation}
The Darmois-Israel gluing conditions \cite{Darmois}, \cite{Israel} demand that along the comoving
boundary the first and second fundamental forms are continuous. In the spherically symmetric case   the radius $R$ has to be continuous and in addition:  i)  the excess mass $m$ should be exactly equal to the mass of  dust in the excised ball; ii)    the boundary should be comoving with the Hubble velocity $H$. This guarantees that  the stress-energy tensor $T_{\mu \nu }$ does  not form the surface layer at the junction surface. Balbinot et al. have found an explicit  solution that describes the corresponding  matching between the  Schwarzschild-de Sitter and FLRW (dust$ +\Lambda $) spacetimes \cite{Balbinot1988}. Therefore we stop discussion on FLRW and SdS (SAdS) solutions at this point and focus on the description of the accretion zone.
  
We assume that in the accretion region the Einstein equations can be approximated by a set of stationary equations  in time intervals that are much smaller than a characteristic time $T $ that is defined in the next section. Let  $R_\infty $ be the    size of the cloud of gas and $a_\infty $   the asymptotic speed of sound. Outside $R_\infty $ the geometry can be connected  smoothly to the SdS or SAdS spacetime geometry by a transient zone of a negligible    mass.  Let us remark that the  assumption of the approximate stationarity  can be checked aposteriori, after finding appropriate solutions. Approximately stationary accretion equations have been derived in \cite{Malec99}.
A quasi-stationary solution, of a similar accretion problem in the asymptotically flat spacetime with a spherical black black hole, appears to  be stable under axially symmetric perturbations \cite{Mach2008}.

\section{Equations of steady accretion}

The metric in the accretion zone is spherically symmetric and  has the form
\begin{equation}
ds^2=-N^2dt^2+\hat adr^2 +R^2\left( d\theta^2 +\sin^2 (\theta )d\phi^2\right) ,
\label{ac}
\end{equation}
where we use comoving coordinates $t, r, 0\le \theta \le \pi , 0\le \phi < 2\pi$ --- time, coordinate radius and two angle variables, respectively. $R$ denotes the areal radius and $N$ is the lapse. $\hat a$ is the radial-radial metric component. The radial velocity of gas is given by $U = \frac{1}{N} \frac{dR}{dt}$.  

The energy-momentum tensor reads   $T_{\mu \nu }^B =\left( \rho +p \right) U_\mu U_\nu +pg_{\mu \nu }$ with the time-like   normalized four-velocity $U_\mu $, $U_\mu U^\mu =-1$.  A comoving observer would measure local mass   density $\rho =T^{\mu \nu }U_\mu U_\nu $.  
Let   $n_\mu$ be the unit normal to a  coordinate sphere lying in the hypersurface $t=const$ and let $k$ be the related mean curvature scalar, $k={R\over 2}\nabla_i n^i=\frac{1}{\sqrt{\hat a}}\partial_rR$.  We assume  the perfect gas equation of state $p=\left( \Gamma -1 \right) \rho_0 \epsilon  $, where $\epsilon $ is the  specific internal energy and $\Gamma $ is a constant. Assuming that the accretion is isentropic, one can derive the
 polytropic equation of state $p=K\rho_0^\Gamma $, with a constant $K$. The  internal energy  $E=\rho_0 \epsilon $ and the rest  $\rho $ and baryonic  $\rho_0$  mass densities are related by $\rho =\rho_0+E= \rho_0+p/(\Gamma -1)$.

The matter-related and geometric quantities satisfy Einstein equations  and the baryonic mass conservation. (But let us point out that for isentropic flows the conservation of baryonic mass also follows from Einstein equations \cite{Malec99}.) One can find the mean curvature $k$  from the Einstein  constraint equations $G_{\mu 0}=8\pi T_{\mu 0}$  \cite{Malec99} 
\begin{eqnarray}
&& k = \sqrt{1-\frac{2m(R)}{R}  -{\Lambda \over 3} R^2+U^2},
\label{mck}
\end{eqnarray}
where   $m(R)$ is the quasilocal mass,  
\begin{equation}
m(R)=m-4\pi \int_R^{R_\infty }dr r^2 \rho   .
\label{localmass}
\end{equation}
In the line element (\ref{ac}) we have the comoving time. In the polar gauge foliation one has a new time $t_S(t,r)$ with $\partial_{t_S}=\partial_t-NU\partial_R$. The field  $\partial_{t_S}$ is tangent to the cylinder of constant areal radius,   $\partial_{t_S}R=0$.

 One can show that 
\begin{equation}
\dot M_\rho \equiv \partial_{t_S} m(R)=
 4\pi NU R^2 \left( \rho +p \right)  .
\label{massconservation}
\end{equation}
The mass contained in an annulus $(R, R_\infty )$ changes if the fluxes on the right hand side, one directed outward and the other inward, do not cancel. 
The baryonic current density reads   $j^\mu \equiv \rho_0 U^\mu$.  Its continuity equation   reads
\begin{equation}
\nabla_\mu j^\mu = 0.
\label{bc}
\end{equation}
For stationary flows the local baryonic flux  
\begin{equation}
\dot M=-4\pi U R^2\rho_0. 
\label{1}
\end{equation}     
is time-independent at a fixed $R$. 

We say that the accretion process is   stationary (or quasi-stationary) if all relevant physically observables, that are  measured at a fixed areal radius $R$, remain approximately constant during time intervals much smaller than the  runaway instability time scale  $T=M/\dot M  $.  That means that  $\partial_{t_S}X\equiv (\partial_t -NU\partial_R) X = 0$ for $X=\rho_0, \rho, j, U\ldots  , \dot M $.

For quasistationary flows we  have  $\partial_R \dot M_{\rho } =0 $  and $\partial_R\dot M = 0$ \cite{Malec99}.  Therefore both $ \dot M_{\rho }$ and $\dot M$ are equal, modulo a constant factor; they can be identified. The quantity $\dot M$ will be called the mass accretion rate.
One obtains from (\ref{1}) an  expression
\begin{equation}
\partial_RU^2= -{4U^2\over R} -2U^2 \partial_R \ln \left( \rho_0 \right)
\label{U2}
\end{equation}
The speed of sound is defined as $ a=\sqrt{\partial_\rho p}$. It is a useful and straightforward exercise to express the hydrodynamic quantities in terms of $a$:
\begin{eqnarray}
&p&=\rho_0 {\Gamma -1\over \Gamma} {a^2\over \Gamma -1-a^2 }, \nonumber\\
&& \rho = \rho_0   {\Gamma -1\over \Gamma -1-a^2 }-p, \nonumber\\
\rho_0 &=&\rho_{0\infty } \left( {a\over a_\infty }
\right)^{2\over \gamma -1}
  \left( {1-{a^2_\infty \over \Gamma -1}
\over 1-{a^2  \over \Gamma -1}} \right)^{1\over \Gamma -1} .
\label{5}
\end{eqnarray}
Here (and below) quantities with the suffix $\infty $ do refer to their asymptotic values (i. e., at $R_\infty $). Notice that $p, \rho_0$ and $a^2$ show the same monotonicity behaviour: $\partial_R\rho_0= C_1\partial_Rp= C_2 \partial_R a^2$, where $C_1, C_2$ are strictly positive functions.
 
There are two conservation equations   that originate from  the contracted Bianchi identities $\nabla_\mu T^{\mu }_\nu =0$. One of them can be eliminated, if we choose instead  the baryonic mass conservation.  The case when $\nu =0$ is the   relativistic version of the Euler equation,   
\begin{equation}
\label{euler}
N{d\over dR}p + \left(p + \rho\right){d\over dR} N = 0.
\end{equation}
One can solve Eq. (\ref{euler}), using Eq.  (\ref{5}):
\begin{equation}
N =\tilde C  \left( \Gamma -1-a^2 \right) .
\label{4}
\end{equation}
The whole  system of algebraic equations (\ref{mck}) --- (\ref{4}) closes with the imposition of the  Einstein equation, $G_{rr}=8\pi T_{rr}$, which is the only integro-differential  equation. It can be put in the following form
\begin{eqnarray}
&&{d\over dR} \ln \left( a^2\right) =- {\Gamma -1-a^2\over a^2 -{U^2\over k^2}}\times \nonumber \\
&&  {1\over k^2R} \left( {m(R)\over R} -2U^2 +4\pi R^2p -{\Lambda R^2 \over 3}
\right)   .
\label{6}
\end{eqnarray}
Finally the line element, in $(t, R)$ coordinates, is given by
\begin{equation}
ds^2 = -\left( N^2-U^2\right) dt^2 -2{N\over k} dtdR +{dR^2\over k^2} +R^2 d\Omega^2.
\end{equation}
We shall study  transonic accretion flows. For them  the principal object of interest is a sonic point,  where  both the denominator $a^2 -{U^2\over k^2}$ and  the numerator ${m(R)\over R} -2U^2 +4\pi R^2p -{\Lambda R^2 \over 3}$ of Eq. (\ref{6}) vanish.    The corresponding value of the areal radius is called the sonic radius, denoted as  $R_*$. A closer inspection of the sonic point shows that it is a critical point, with branching pairs of solutions describing accretion or wind. In the accretion branch, below the sonic  point the infall velocity $|U|/k $ is bigger than $a$, while outside the sonic sphere the converse is true.   This analysis is very much standard, replicating the work done in the Newtonian case by Bondi \cite{Bondi} and in the general-relativistic case by Malec \cite{Malec99}.  

\section{Qualitative analysis of solutions: SdS black holes }

In this section we shall investigate some properties  of transsonic solutions for  positive values of $\Lambda $.  The speed of sound $a$ (and thus also $\rho_0$ and $p$)   decreases as a function of $R$  --- for accreting solutions ---  when the cosmological constant is absent. We shall show, In Lemma 2,  that for large values of $\Lambda R^2$ the converse is possible.

{\bf Lemma 1. }

  Assume  $\Lambda <  {3m\over 2R^3_\infty  }   $.   Then 
the speed of sound, the pressure  and the  mass   density $\rho_0$  are decreasing outside the sonic sphere, $ {d\over dR}X <0$ for $X=a^2, p  $ and $\rho_0 $.

{\bf Proof of Lemma 1.}

Let us define $C(R) \equiv {m(R)\over R } -2U^2(R)  +4\pi R^2 p -{\Lambda R^2\over 3} $ and let $C_\infty \equiv C(R_\infty )$. One easily obtains from this definition, using equation (\ref{U2}) in order to eliminate $\partial_R U^2$, that  
\begin{eqnarray}
{d\over dR} C(R) &=& {-4C\over R} +4\pi R \left( \rho +6p-{\Lambda \over 2\pi }\right) + \nonumber\\
&&  {3m\over R^2}+4U^2{{d\over dR}\rho_0\over \rho_0} +4\pi R^2{d\over dR}p.
\label{C_equation}
\end{eqnarray}
The condition $\Lambda <  {3m\over R^3_\infty }  $  guarantees that $C_\infty >0$ and --- using the argument of continuity ---  the positivity of the function $C(R)$ in an open interval  $\left( R_s,R_\infty \right)$. We show that $R_s=R_*$.

Notice that while at the sonic point  $C(R_*)=0$,  the function $\partial_R C$ must be strictly positive.  Indeed, assume the opposite;  from continuity $C$ would have to be negative just above $R_*$. Then from Eq. (\ref{C_equation}) it  would follow that both terms with derivatives must be negative, that is --- see the  remark below  (\ref{5}) ---  $\partial_R a^2 <0$.   But  Eq. (\ref{6}) gives $\partial_Ra^2>0$, if $C<0$ and $R>R_*$. This contradiction proves our statement and implies the positivity of $C$  in an  interval $\left( R_*, R_b \right) $ .  

 Now, assume that $C$ changes sign at $R_b$,   that is   $C\left( R_b\right) =0$ and  ${d\over dR}C\left( R\right)_{R=R_b} <0$. But if  $C(R_b)=0$ then --- from Eq. (\ref{6}) ---  the derivative $\partial_Ra^2$ would have to vanish. From (\ref{5}) would then follow  $\partial_R\rho_0=\partial_Rp=0$.   Thus Eq. (\ref{C_equation}) yields at $R_b$
\begin{equation}
{d\over dR} C =    4\pi R_b \left( \rho \left( R_b\right)+6p\left( R_b\right)-{\Lambda \over 2\pi }\right) +  {3m\over R^2_b};
\label{grad_C}
\end{equation}
but this implies ${d\over dR}C>0$ at $R_b$, due to the condition  $\Lambda < {3m\over R^3_b} $. This contradiction proves that $C(R)$ is strictly positive outside the sonic sphere.

The positivity of $C$ allows one to conclude --- from Eq. (\ref{6}) --- that $\partial_R a^2 $ is strictly decreasing outside the sonic sphere. This in turn implies the decrease of the  pressure $p$ and the baryonic mass density $\rho_0$  (see the remark under Eq. (\ref{5})).

 {\bf Lemma 2.}   

  Assume  $\Lambda  R^2_\infty > {18 \alpha a^2_\infty \over \Gamma -1-a^2_\infty   }   $ and define $R_0 = \left(  { 6m   \over \Lambda  }\right)^{1\over 3}$.   The  necessary condition for the existence of solutions is   $\alpha \le 1 $. If $\alpha >1/3$ then   $R_0\ll R_\infty $ and   the speed of sound, the pressure  and the  mass   densities  are increasing   in the interval $\left( R_0, R_\infty \right) $:

 $ {d\over dR}X >0$ for $X=a^2, p$ and $ \rho_0 $.

{\bf Proof of the Lemma 2}.

 The estimate $R_0\ll R_\infty $ follows immediately from the boundary condition $a^2_\infty \gg m/R_\infty $ and the assumed value  of $\Lambda $. In the asymptotic end  $\left( R_0, R_\infty \right) $   the function $-C(R)$ can be bounded from below as follows,
\begin{eqnarray}
-C\left( R\right) &=& {-m(R)\over R } +2U^2(R)  -4\pi R^2 p +{\Lambda R^2\over 3} \ge 
\nonumber\\
&& {-2m\over R }    +{\Lambda R^2\over 3}.
\label{estimate_lemma2}
\end{eqnarray} 
The second line follows from $U^2\ge 0$ and from  the observation that $p\le \rho $ --- thus  $4\pi R^2 p\le 4\pi \int _R^{R_\infty } dr r \rho$ ---  and $4\pi \int _R^{R_\infty } dr r \rho \le {m\over R}$.

Taking into account the last estimate, we can estimate from below the right hand side of Eq. (\ref{6})  by 
\begin{equation}
D(R)\equiv   {\Gamma -1-a^2_\infty \over a^2  } 
  \left( -{2m\over R^2}    + {\Lambda R \over 3}\right) .
\label{D}
\end{equation}
We obtain from Eqs (\ref{6}) and (\ref{D}) the inequality
\begin{equation}
{d\over dR}a^2 \ge \left( \Gamma -1-a^2_\infty \right) 
  \left( -{2m\over R^2}    + {\Lambda R \over 3}\right) .
\label{D_inequality}
\end{equation}
It is clear that i) $R_0$ is the  null point of $D(R)$; ii)  the function  $D(R)$ is strictly positive for 
$R\in \left( R_0, R_\infty \right) $. Therefore $a^2$ (and consequently the remaining gas characteristics $p$ and $\rho_0$ ) does increase in this interval, provided that a solution does exist.

The necessary condition for the existence of a solution is that $a(R_0)$ should be  strictly positive. Integration of Eq. (\ref{D_inequality}), between $R_0$ and $R_\infty $, yields 
\begin{eqnarray}
&& a^2_\infty - a^2\left(R_0\right)  \ge \left( \Gamma -1-a^2_\infty \right) \times \nonumber\\
&&
  \left( -{2m\over R_0}    - {\Lambda R^2_0 \over 6} +{2m\over R_\infty }    + {\Lambda R^2_\infty \over 6}\right) .
\label{a2R_inequality}
\end{eqnarray}
We can replace ${2m\over R_0}$ by     $ {\Lambda R^2_0 \over 3}$, using the  definition of $R_0$, and get the following 
\begin{eqnarray}
    &&a^2\left(R_0\right)  \le  \left( \Gamma -1-a^2_\infty \right) 
   {\Lambda R^2_\infty  \over 18} \left(       {6R_0^2 \over  R_\infty^2}  -1 \right) +a^2_\infty  - 
\nonumber\\ 
&& \left( \Gamma -1-a^2_\infty \right) {2m\over R_\infty } \le a^2_\infty + \nonumber\\
&& \left( \Gamma -1-a^2_\infty \right) 
   {\Lambda R^2_\infty  \over 18} \left(  {6 \times 36^{1\over 3}\left( {m \over R_\infty } \right)^{2\over 3}  \over  \left(  \Lambda R_\infty^2\right)^{2\over 3}}  -1 \right)   .
\label{a2R_final}
\end{eqnarray}
In the second inequality we dropped the positive term with $m/R_\infty $ and used the definition od $R_0$.
We already assumed that $m/R_\infty \ll a^2_\infty $which ensures that the first term within second bracket is small and the right hand side of the last inequality in (\ref{a2R_final}) is approximated by  $E(R)\equiv  a^2_\infty - {\Lambda R^2_\infty  \over 18}  \left( \Gamma -1-a^2_\infty \right)   $. Therefore   if the cosmological constant  is given by $\Lambda R^2_\infty =18 \alpha {a^2_\infty \over     \Gamma -1-a^2_\infty }  $, with $\alpha $       larger than $1$, then $E(R)$ must be negative and  the square of the speed of sound at $R_0$ must be nonpositive , which means that solutions are absent. On the other hand solutions can exist for $0.01< \alpha <1$;  in this case $C(R)<0$ between $\left( R_0, R_\infty \right) $, which implies that  $ {d\over dR}X <0$ for $X=a^2, p, \rho_0 $ and $\rho $. 

The two lemmas proven in this section have a transparent physical interpretation. Lemma 1 assumes   that the dark energy density $\Lambda /4\pi $ is smaller than one half of    the averaged matter density $3m/4\pi R^3_\infty $. The presence of $\Lambda $ is expected to have some quantitative impact onto accretion mass rate $\dot M$, but    qualitative features of the flow  are not influenced   --- in particular the gas characteristics $a$, $|U|, \rho_0$ and $\rho $ are all decreasing functions.

Lemma 2 assumes the opposite --- that the dark energy density $\Lambda /4\pi $ is much larger than   the average matter density $3m/4\pi R^3_\infty $.  Above a particular value --- that depends on the boundary characteristics of the flow, its volume and the black hole  mass --- steadily accreting solutions do not exist. This is intuitively understandable  --- large $\Lambda $ implies a large Hubble expansion velocity  which can obstruct  or even prohibit accretion, through the junction condition on the boundary of the vacuole. Therefore one can expect a significantly smaller mass accretion rate $\dot M$ and even its absence.

\section{Qualitative analysis of solutions: SA{\small d}S black holes }

{\bf Lemma 3. }  Assume $\Lambda <0$.   Then:

i)  for $R>R_*$ 
the speed of sound, the pressure  and the  mass   densities $\rho_0$ and $\rho $  are decreasing,  
 $ {d\over dR}X <0$ for $X=a^2, p$ and $ \rho_0 $.

ii)  Solutions are absent if $\Lambda R^2_\infty \le  -6{ \Gamma -1 -a^2_\infty \over   \Gamma -1-a^2_*} $.

{\bf Proof.} 

Part i) The key observation is that the function $C$ is strictly positive, if $\Lambda <0$.  Indeed,  let $R_b>R_*$ be a smallest radius at which $C(R_b)$ vanishes, and  assume $C<0$  in the open interval $\left( R_*, R_b\right) $ (notice that  $C\left( R_* \right) $ vanishes). But  Eq. (\ref{C_equation})  would imply in such a case  that both terms with derivatives become negative in an sub-interval $\left( R_*, R_1\right) $ of  $\left( R_*, R_b\right) $, that is --- see the  remark below  (\ref{5}) ---  $\partial_R a^2 <0$. But this in turn would   contradict    Eq. (\ref{6}) ---  if $C<0$  then  $\partial_Ra^2>0$ outside the sonic sphere. Thus we arrive to the  contradiction  in this sub-interval  $\left( R_*, R_1\right) $ and  the argument of continuity and the definition of $R_b$ imply the positivity of $C$  in  the open interval $\left( R_*,R_b\right) $.  

In the next step we show that $C$ cannot vanish outside the sonic sphere. Indeed,    (\ref{6}) and  (\ref{C_equation}) imply that ${dC\over dR}$ is strictly positive whenever $C=0$ for $R>R_*$; that in turn  yields (by the argument of continuity) the strict positivity of $C$ outside the sonic sphere. Eq. (\ref{6}) allows one to  conclude that the speed of sound is monotonically decreasing, and that implies also decrease of the mass density $\rho_0$
 and the pressure.  

Part ii) In order to prove  the necessary existence condition, notice that  Eq. (\ref{6}) gives, after integration, 
\begin{equation}
|a^2_\infty - a^2(R)|\ge \int_R^{R_\infty }dr | \left( \Gamma -1-a^2  \right) {1\over r} \left(   -2U^2   -{\Lambda r^2 \over 3}\right) |.
\label{Lemma3}
\end{equation}
Notice that $a^2<\Gamma -1$; thus $ |a^2_\infty - a^2(R)| \le \Gamma -1-a^2_\infty $. The conservation of 
the mass accretion rate $\dot M$  and  ${d\over dR }\rho_0\le 0 $ lead to the estimate ${d\over dR} \left( U^2R^4 \right) \ge 0$, and this in turn implies inequality 
\begin{equation}
U^2\le U^2_\infty {R^4_\infty \over R^4 }.
\label{Lemma3_a}
\end{equation}
Collecting this information, one finally obtains
\begin{equation}
\Gamma -1 -a^2_\infty \ge \left( \Gamma -1-a^2(R)  \right)   \left(   -{U^2_\infty \over 2}  
 -{\Lambda R^2_\infty  \over 6} \left(  1-  {R^2\over R^2_\infty }\right)  \right) .
\label{Lemma3_b}
\end{equation}
Let $R=R_* $ and ${\Lambda R^2_\infty  \over 6}<-{ \Gamma -1 -a^2_\infty \over    \Gamma -1-a^2_*}$  ; then --- taking into account that  ${R^2\over R^2_\infty }$ is negligibly small, and the same is true for  the velocity term $U^2$--- we conclude that the right hand side of  Eq. (\ref{Lemma3}) exceeds its left hand side. This contradiction proves the necessary condition
of Lemma 3.

\section{Description of numerics}

Our aim is to find transsonic accretion flows.   We assume the total mass $m=1$, $\Gamma =4/3$ and $R_\infty =10^6$. The boundary values  at $R_\infty $ are $k=N=\sqrt{1-{2m\over R_\infty } -U^2_\infty -{\Lambda \over 3} R^2_\infty }$  and  $a^2_\infty =2\times 10^{-4}$.  There are three remaining parameters that are free: the cosmological constant $\Lambda $,  the mass accretion rate $\dot M$ and the asymptotic density $\rho_{0\infty }$ --- after specifying them, equations are integrated inward, starting from $R_\infty $. These quantities  should be such as to ensure the asymptotic conditions: $|U_\infty |\ll \sqrt{2m/R_\infty }\ll  a_\infty $.

  A numerical run, with the above boundary  data,  has been  regarded as   succesful, if:

i) the integration  started  from $R_\infty $ and reached  the outermost  apparent horizon, that the  point $R_i$ such that ${2m(R_i)\over R_i}+{\Lambda R^2_i \over 3} =1$,  along the accretion branch (with the increasing--- in the direction to the centre --- infall velocity $|U|$);

ii)  there appears  a  sonic point  at a radius $R_* \ge R_i$. 

More specifically, in order to find a single transsonic flow, we fix $\Lambda $ and the mass accretion rate  $\dot M$,  
and treat the asymptotic value of the baryonic mass density  as a free parameter. For a random choice of $\rho_{0\infty }$ there are three possibilities, one exceptional and two generic:

a)  an almost unlikely event, that the numerical run is succesful in the sense defined above and we find the sought accretion flow;

b) the solution numerically fails --- a singularity appears during integration;

c) there exists a numerical solution without a sonic point --- a piece of the solution describes  accelerating  accreting flow, while the other corresponds to a deccelerating accreting gas.

In the case  a) the job is done. If one of the remaining  generic alternatives occurs --- say b), with the asymptotic density $\rho_{b)0\infty }$ --- we change $\rho_{0\infty }$ until we get the situation described in c) with the asymptotic density $\rho_{c)0\infty }$.  Now  we use the bisection method in order to find an intermediate value $\rho_{i0\infty }$ between the two values  $\rho_{b)0\infty }$ and  $\rho_{c)0\infty }$, for which the numerical run  is succesful and the transsonic flow exists.   

The numerics itself is standard --- we use commonly known algorithms of  the 8th order.

 The calculations  had been performed for a few dozens of values --- negative and positive ---  of the  cosmological constant $\Lambda $.  
In the  sector of negative cosmological constant  we take $\Lambda \in \left(  -5\times 10^{-13}, -10^{-19} \right) $; the absolute value over bound is larger by one order of magnitude than  the value found in Lemma 3 as a necessary existence  condition.  In the positive part of the spectrum we choose  $\Lambda \in \left(  10^{-19}, 7.2 \times 10^{-15}  \right) $; this is quite close to the  value established  as the necessary condition in Lemma 2. For each fixed  value of $\Lambda $ we find about 100 solutions corresponding to different values of $\dot M$.

The  obtained information is summarized in figures 1 and 2. The abscissa shows the ratio $x=M_g/m$ of the mass $M_g=4\pi \int_{R_i}^{R_\infty }dr r^2 \rho $  of gas to the total mass $m$. The total mass has been normalized to 1. We obtain, although in a less transparent form, the same result as in  (\cite{Malec2006}) --- the maximum of the mass accretion rate corresponds to $x\approx 1/3$. For a fixed value of $\dot M$ there exist two flows, one with heavier center and lighter gas, and the other with opposite characteristics. Just around the maximum of $\dot M$ there are blank points --- we find that in this region it is difficult to find a numerical solution. We think that this a purely numerical artefact.   There is a slight shift of this maximum towards smaller values of $x$ with the increase of the absolute value of the  cosmological constant.

\begin{figure}[h]
\begin{center}
\includegraphics[width=\linewidth]{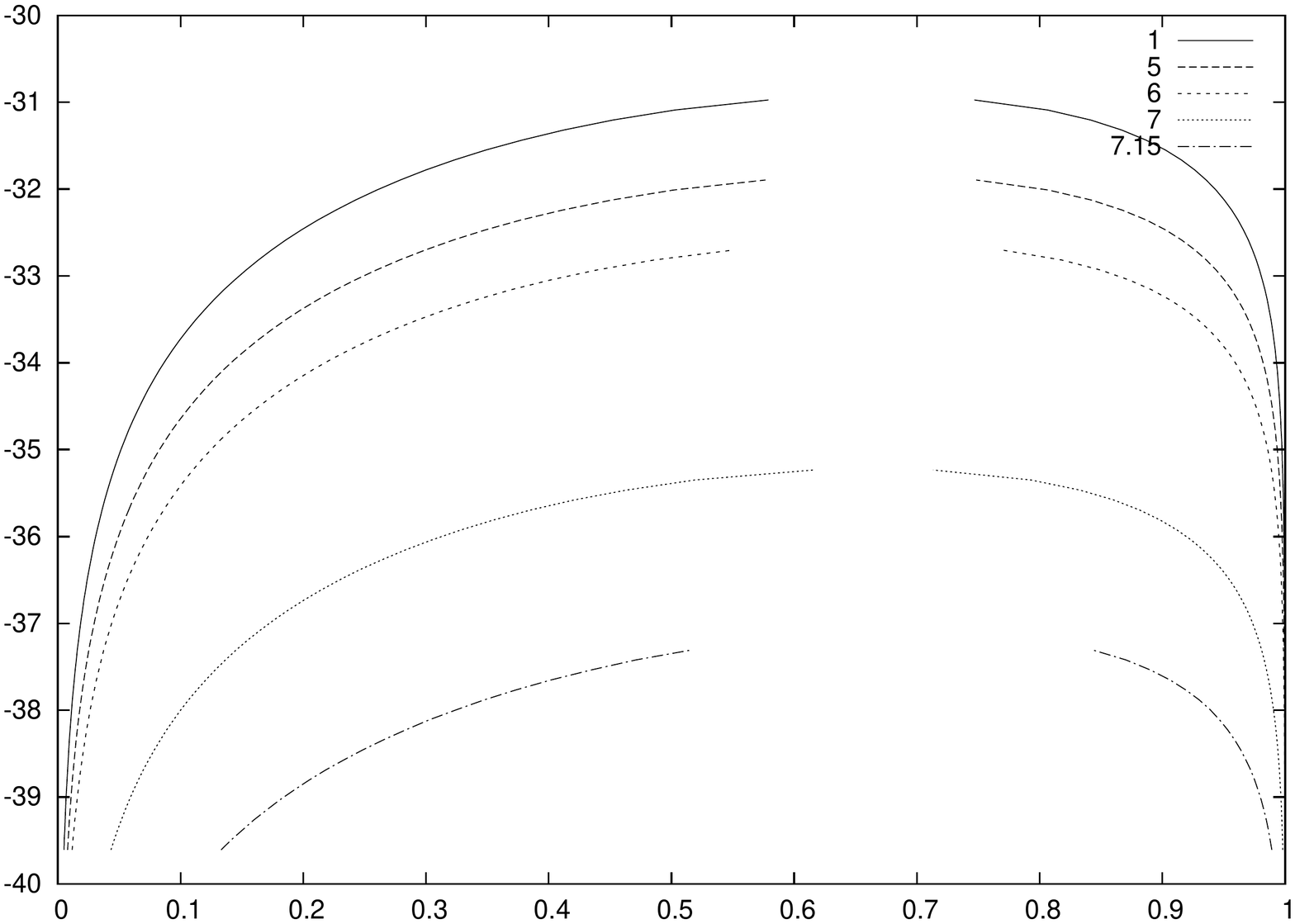}
\end{center}
\caption{\label{fig:1}   The ordinate shows the mass accretion rate  $\dot M$  and the abscissa shows  $1-x $, where $x$ is the relative  mass  of gas in the system.. The  various lines  correspond  to $\Lambda R^2_\infty  /10^{-3} =1,  5, 6, 7, 7.15$ in the    order from the top.} 
\end{figure}

\begin{figure}[h]
\begin{center}
\includegraphics[width=\linewidth]{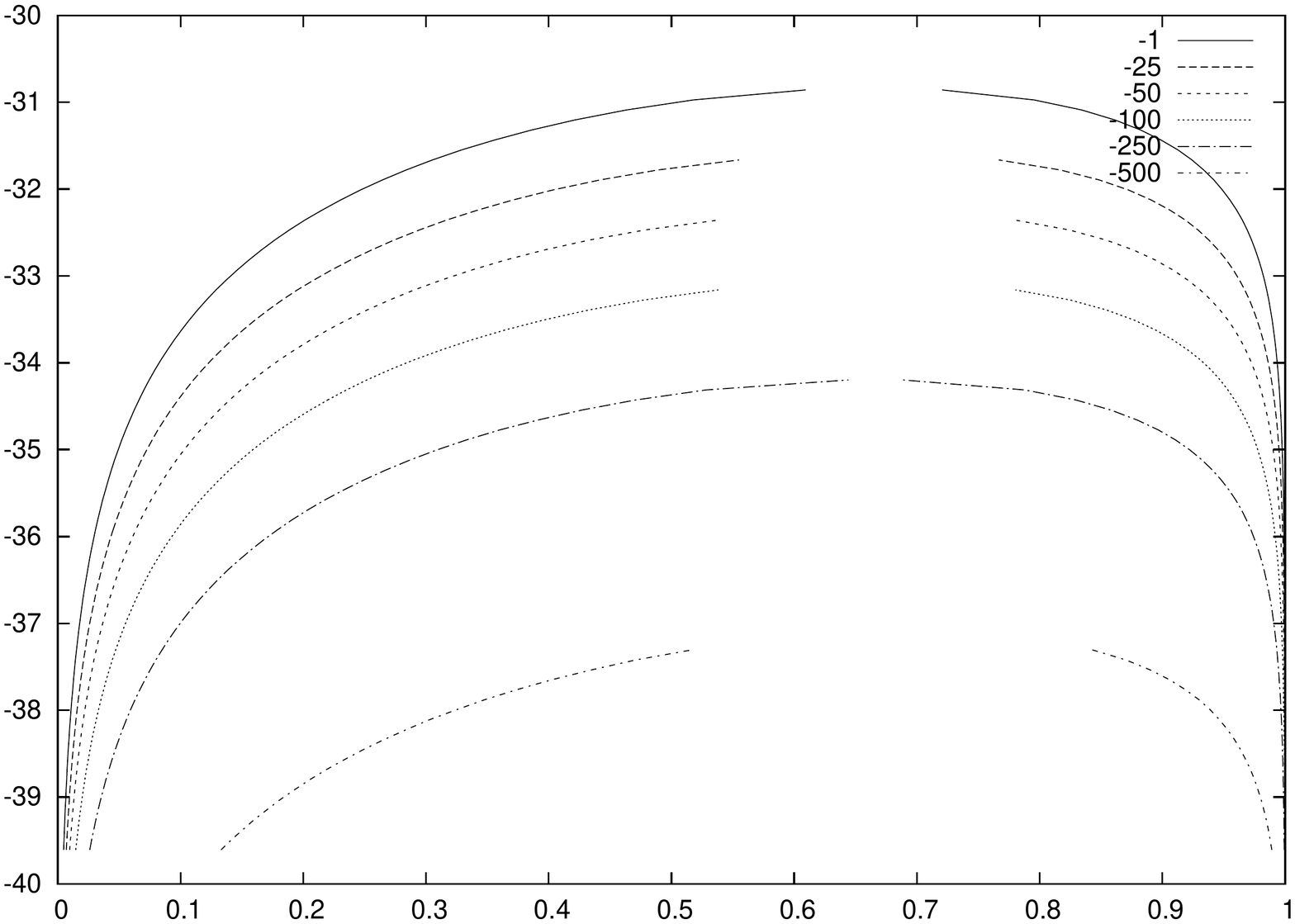}
\end{center}
\caption{\label{fig:2} The ordinate shows the mass accretion rate  $\dot M$  and the abscissa shows $1-x $, where $x$ is the relative  mass  of gas in the system. The  various lines  correspond  to $\Lambda R^2_\infty  /10^{-3} =-1,  -25, -50, -100, -250, -500$ in the decreasing order from the top.  }
\end{figure}

The only new  relevant physically information is that concerning the mass accretion rate $\dot M$.
It appears that the presence of the cosmological constant manifests only when the dark energy density $\Lambda /(4\pi )$ exceeds the averaged matter density $3m/(4\pi R^3_\infty ) \approx 4\times 10^{-19}$ by at least three orders of magnitude.  The mass accretion rate $\dot M$  is decreasing with the increase of the absolute value of the cosmological constant. It is clear that  the rate of the falloff depends on the sign of $\Lambda $  ---  it is faster for positive values  ----  but the falloff itself occurs both for positive and negative cosmological constant.

\section{Cosmological implications}

Results that are reported in the preceding section demonstrate that while  the mass accretion rate  depends on   the cosmological constant, this effect becomes significant only when the dark energy  fraction $\Omega_\Lambda $ is much larger than  the material fraction $\Omega_m$ .  On the other hand, this impact can be dramatic,  a sevenfold  increase of  $\Lambda $  --- starting from its value corresponding   to $\Omega_\Lambda /\Omega_m \approx 10^3$ --- leads to the diminishing 
of $\dot M$ by seven orders in magnitude, as seen in  Fig. 1. This is so for the accretion onto SdS black hole, but the phenomenon is strong also in the case of SAdS (Fig. 2) .
When applying these results  to our Universe, we have to take into account, that $\Omega_\Lambda \gg \Omega_m$ in two epochs --- during  inflation and after $10^{11}$ years from Big Bang.

Inflation  strictly excludes steady accretion onto primordial black holes, due to Lemmas 2 and 3,  because during  inflation era    $\Omega _\Lambda $ exceeds $\Omega_m $ by something like 50 orders of  magnitude.  This fact is concordant  with the well known freezing of structures that are bigger than a particle horizon \cite{Mukhanov2005}, but the difference is that now it applies to small accretion systems that are located well within the horizon.  

In the present Universe  the dark energy and material densities are roughly the same, which suggests that the effect of dark energy is negligible. The mass density of largest bound structures - galactic superclusters - exceeds the cosmological  mass density by one order of magnitude.  Notice, however, that   the tenfold ageing of the Universe would increase the ratio $\Omega_\Lambda /\Omega_m$ by a factor of 100. That means that in the Universe older than $10^{11}$ years the steady accretion would become less efficient and at the time  $t\rightarrow 10^{12} $ years its efficiency goes to zero. 

Penrose conjectured  the so-called “Weyl curvature hypothesis” \cite{Penrose2004}.  It asserts, in its informal version,  that     a Friedman-Lemaitre-Robertson-Walker spacetime, where the   Ricci curvature is nonzero but the   Weyl curvature vanishes, evolves towards a vacuum spacetime filled with black holes 
and gravitational radiation ---  with nonvanishing Weyl curvature and  negligible Ricci tensor. (A more formal scenario has been worked out by Tod \cite{Tod}). This picture assumes that initially small inhomogeneities of FLRW universe   accrete matter and transform themselves  into black holes, that   gradually merge, leaving at the end a net of huge  black holes and a lot of gravitational radiation.  A cautious intepretation  of our findings would be to say that  the role of steady accretion  in the realization of this scenario, in the presence of dark energy, is insignificant.  It might well happen, however, that this property of dark energy of diminishing the steady spherical accretion signals a more general feature --- that dark energy damps any accreting process. In such a case there arises another  fundamental question: how dark energy impacts this  scenario outlined in the   “Weyl curvature hypothesis”?

 Acknowledgements. This research was carried out with the supercomputer
“Deszno” purchased thanks to the financial support of the European Regional
Development Fund in the framework of the Polish Innovation Economy
Operational Program (contract no. POIG. 02.01.00-12-023/08).

\end{document}